\documentclass[aps,prd,preprint,groupedaddress,showpacs]{revtex4}
\usepackage[dvips]{graphicx}
\begin{document}
\preprint{KOBE-TH-02-03}
\preprint{UT-02-63}
\title{Stable Neutral Fermi Ball}
\author{K.Ogure}
\affiliation{Department of Physics, Kobe University, Rokkoudaicho 1-1, Nada-Ku, Kobe 657-8501, Japan}
\author{T.Yoshida}
\affiliation{Department of Physics, Tokyo University, Hongo 7-3-1, Bunkyo-Ku, Tokyo 113-0033, Japan}
\author{J.Arafune}
\affiliation{National Institution for Academic Degrees, Hitotsubashi 2-1-2, Chiyoda-Ku, Tokyo 101-8438, Japan}
\date{\today}
\begin{abstract}
Fermi Ball is a kind of nontopological soliton with fermions trapped in its domain wall, and is suggested to arises from the spontaneous symmetry breaking of the approximate $Z_2$ symmetry in the early universe. We find that the neutral thin-wall Fermi Ball is stable in the limited region of the scalar self-coupling constant $\lambda$ and the Yukawa coupling constant $G$. We find that the Fermi Ball is stabilized due to the curvature effect of the domain wall caused by the fermion sector. We also discuss whether such stable Fermi Balls may contribute to the cold dark matter. 
\end{abstract}
\pacs{05.45.Yv, 95.35.+d}
\maketitle

\section{INTRODUCTION\label{intro.sec}}
In quantum field theory and especially in the cosmological context, various models have been discussed where the spontaneous breaking of a discrete symmetry produces domain walls \cite{KIB,VIL}. If the symmetry is approximate and broken spontaneously in the history of the expanding universe, the false vacuum regions shrink due to the energy density difference \cite{GEL} and this process is accelerated by the surface tension.  The shrinking will stop before collapsing if there exist zero mode solutions for the fermions bound in such domain walls \cite{DVA,SAK} and the Fermi pressure of such fermions becomes comparable to the shrinking force due to the surface tension and the volume energy.  These objects, a kind of non-topological solitons, are called Fermi Balls \cite{MAC} and introduced as a candidate for cold dark matter \cite{MAC,MACC}.  They were also suggested in the baryon-separation scenario with the QCD energy scale \cite{BRA}. However, the stability of the Fermi Balls which should be essential to be a candidate for cold dark matter has not been fully examined.

We consider in this paper the stability of the thin-wall Fermi Balls against the fragmentation. The Fermi Ball, if it is electrically neutral as was firstly proposed \cite{MAC}, is unstable against the fragmentation, when the domain wall thickness $\delta_b$ can be neglected in comparison with the Fermi Ball radius $R$.  In the previous paper \cite{YOS}, we estimated the corrections caused by the finite $\delta_b$ effect (corresponding to the curvature effect of the domain wall), and found that the neutral Fermi Ball is unstable even if the curvature effect is included. Though the electrically charged Fermi Ball introduced by Morris \cite{MOR} has the stability against the deformation from the spherical shape \cite{MOR,YOS}, they are metastable \cite{YOS} and not absolutely stable against the fragmentation. We do not discuss the charged Fermi Ball in the present paper.

In the above discussions,  though, the thickness $\delta_f$ of the distribution of the fermions confined in the wall is much less than the wall thickness $\delta_b$ and can be neglected. This approximation is valid only for the case where the fermions are tightly bound to the domain wall. We in this paper consider the case where $\delta_f$ cannot be neglected. In order to estimate the corrections caused by the finite $\delta_f$ effect, we use the perturbation approximation expanding the scalar field $\phi$, the fermion field $\psi$ and the static energy of the Fermi Ball $E$ in the power of $\delta_b/R$.  We examine the stability of the Fermi Ball against the fragmentation at each level of the perturbation.

The contents of the present paper are organized as follows:  We first explain the neutral Fermi Ball model to clarify terminologies and method of energy estimation in Sec.\ref{setup.sec}. We next examine the stability of the Fermi Ball within the leading (zeroth) order of $\delta_b/R$-perturbation but with finite $\delta_f$ in Sec.\ref{1storder.sec}, and regain the result obtained in the previous works in the limit $\delta_f \rightarrow 0$ \cite{MAC}. The effect of the finite $\delta_f$ in the higher order corrections of $\delta_b/R$-perturbation is investigated in Sec.\ref{2ndorder.sec}. We discuss constraints on the Fermi Ball parameters from the experimental viewpoint in Sec.\ref{para.sec}. We summarize the obtained results in Sec.\ref{concl.sec}.

\section{METHOD FOR ESTIMATING FERMI BALL ENERGY\label{setup.sec}}
We consider the simple model with the Lagrangian density, 
\begin{equation}
{\cal L}=\frac{1}{2}(\partial_{\mu}\phi)^2 
	+\overline{\Psi}(i\gamma^{\mu}\partial_{\mu}-G\phi)\Psi 
	-U(\phi)~~,
\label{L.eqn2}
\end{equation}
where the scalar potential $U(\phi)$ is approximately double-well shaped,
\begin{equation}
U(\phi)=\frac{\lambda}{8}(\phi^2-v^2)^2 +\Delta(\phi) ~~.
\label{U.eqn2}
\end{equation}
Here, the first term has the $Z_2$ symmetry under $\phi \leftrightarrow -\phi$, and the second term violates the symmetry though it is assumed to be much smaller than the first one, $\Lambda \simeq |\Delta(v)-\Delta(-v)| \ll \lambda v^4$. Supposing $\Delta(-v) > \Delta(v)$, we call the region with $\phi=v$ and that with $\phi=-v$ as the true vacuum and the false vacuum, respectively.  In the following, we neglect the second term except for the explicit discussion in Sec.\ref{concl.sec}. Fermi Ball is the ground state of the system with the total number of the fermion being fixed:
\begin{equation}
N_f=\int{\rm d}^3x~\Psi^{\dagger}\Psi~~.
\label{Nf.eqn2}
\end{equation}
The classical fields $\phi(\vec{x},t)$ and $\Psi(\vec{x},t)$ for Fermi Ball extremize 
\begin{equation}
L[\phi,\Psi;\epsilon_f]=\int{\rm d}^3x~{\cal L} 
	+\epsilon_f \left(
		\int{\rm d}^3x~\Psi^{\dagger}\Psi -N_f \right)~~,
\label{Le.eqn2}
\end{equation}
with $\epsilon_f$ the Lagrange multiplier. The static fields thus satisfy
\begin{eqnarray}
&&\left( \vec{\alpha}\vec{p} +G\phi\beta \right)\Psi 
	=\epsilon_f\Psi~~,
\label{PsiEq.eqn2}\\
&&-\vec{\nabla}^2\phi +G\Psi^{\dagger}\beta\Psi 
+\frac{\lambda}{2}\phi(\phi^2-v^2) 
	=0~~,
\label{phiEq.eqn2}
\end{eqnarray}
where $\vec{\alpha}=\gamma^0\vec{\gamma}$, $\beta=\gamma^0$ and $\vec{p}=-i\vec{\nabla}$.

Assuming the spherical symmetry of $\phi({\vec x})$, we take it as a function of the radial coordinate $r$. We first consider Eq.(\ref{PsiEq.eqn2}).  Let $\Psi(\vec{x})$ be the eigenfunction of ${\vec {\bf J}}^2$, ${\bf J_z}$ and ${\bf P}$: 
\begin{equation}
{\bf {\vec J}}^2\Psi_{J}^{M}=J(J+1)\Psi_{J}^{M},\hspace{1cm} 
	{\bf J_z}\Psi_{J}^{M}=M\Psi_{J}^{M},\hspace{1cm} 
	{\bf P}\Psi_{J}^{M}=\beta \Psi(-{\vec x})=P \Psi_{J}^{M}~~.
\end{equation}
Then, $\Psi_{J}^{M}$ is written as
\begin{eqnarray}
\Psi_{J}^{M}(\vec{x})=\frac{1}{r} \left(
	\begin{array}{l}
	f(r){\cal Y}_{lJ}^{M}(\theta,\varphi) \\
	g(r){\cal Y}_{l'J}^{M}(\theta,\varphi)
	\end{array}
\right)~~,
\label{fYgY.eqn2}
\end{eqnarray}
where ${\cal Y}_{lJ}^{M}(\theta,\varphi)$ and ${\cal Y}_{l'J}^{M}(\theta,\varphi)$ are the spherical spinors having the eigenvalues $J=l-\omega/2=l'+\omega/2$ and $M$, with $\omega=\pm 1$.  We take ${\cal Y}_{l'J}^{M}=({\vec \sigma}{\vec x}/r){\cal Y}_{lJ}^{M}$ and note $P=(-1)^l=(-1)^{J+\omega/2}$. 
Substituting Eq.(\ref{fYgY.eqn2}) into Eq.(\ref{PsiEq.eqn2}), we get
\begin{eqnarray}
\left\{
	\begin{array}{l}
	G\phi f +\left( p_r -i\frac{K}{r} \right)g 
	= \epsilon_f f \\
	\left( p_r +i\frac{K}{r} \right)f -G\phi g 
	= \epsilon_f g~~, 
	\end{array}
\right.
\label{fgEq.eqn2}
\end{eqnarray}
where $p_r=-i({\rm d}/{\rm d}r)$ and $K=\omega (J+1/2)$. The equation (\ref{fgEq.eqn2}) is compactly written in terms of $\psi(r)\equiv{f\choose g}$:
\begin{equation}
H_f\psi=\epsilon_f\psi~~,
\label{Hfpsi.eqn2}
\end{equation}
with
\begin{equation}
H_f=\sigma_1p_r +\sigma_2\frac{K}{r} +\sigma_3G\phi~~.
\label{Hf.eqn2}
\end{equation}
The radial coordinate $r$ is hereafter replaced by $w=r-R$.

We next consider Eq.(\ref{phiEq.eqn2}). Noting $\bar{\Psi}\Psi=\sum_{KM}\bar{\Psi}_M^J\Psi_M^J=(1/4\pi r^2)\sum_K (2|K|)\psi^{\dagger}\sigma_3\psi $, we have
\begin{equation}
\frac{{\rm d}^2\phi}{{\rm d}w^2} 
	+\frac{2}{R+w}\frac{{\rm d}\phi}{{\rm d}w}
	=
	\frac{\lambda}{2}\phi(\phi^2-v^2) 
	+\frac{G}{4\pi(R+w)^2}\sum_{K}(2|K|)\psi^{\dagger}\sigma_3\psi~~.
\label{phiEqqq.eqn2}
\end{equation}
The energy of the Fermi Ball is expressed in terms of $\phi$ and $\psi$ as follows:
\begin{equation}
E=E_f+E_b~~,
\label{Etot.eqn2}
\end{equation}
where $E_f$ is the fermi energy,
\begin{eqnarray}
E_f&=&\int{\rm d}^3x~ \Psi^{\dagger}\left( 
	\vec{\alpha}\vec{p} +G\phi\beta \right)\Psi
	\nonumber\\
	&=&\sum_{KM}\int_{-\infty}^{+\infty}{\rm d}w~ 
		\psi^{\dagger}H_f\psi~~,
\label{Ef.eqn2}
\end{eqnarray}
and $E_b$ is the surface energy \footnote{We call $E_b$ the surface energy instead of the boson energy. It is simply the usual terminology which has been used in the previous works concerning Fermi Balls \cite{MAC,YOS,MOR}.},
\begin{eqnarray}
E_b&=&\int{\rm d}^3x~ \left\{ \frac{1}{2}(\vec{\nabla}\phi)^2 
	+\frac{\lambda}{8}(\phi^2-v^2)^2 \right\}
	\nonumber\\
	&=&4\pi R^2\int_{-\infty}^{+\infty}{\rm d}w~ 
		\left( 1+\frac{w}{R} \right)^2 \left\{ 
			\frac{1}{2}\left( \frac{{\rm d}\phi}{{\rm d}w} \right)^2 
			+\frac{\lambda}{8}(\phi^2-v^2)^2 \right\}~~.
\label{Eb.eqn2}
\end{eqnarray}
Note that we estimate the energy by integrating the above integrands not from $-R$ but from $-\infty$, since most of the contribution comes from the region near the surface and the error due to this approximation is exponentially small ($\propto e^{-{\rm Const}. vR}$). 

\section{LEADING ORDER OF $\delta_b/R$-PERTURBATION\label{1storder.sec}}
Let us expand the fields,
\begin{eqnarray}
\left\{
	\begin{array}{l}
	\phi=\phi_0+\phi_1+\cdots  \\
	\psi=\psi_0+\psi_1+\cdots~~, 
	\end{array}
\right.
\label{funcexp.eqn3}
\end{eqnarray}
and the Hamiltonian,
\begin{eqnarray}
H_f&=&H_0+H_1+H_2+\cdots~~, \label{Hfexp.eqn3}\\
	&&H_0=\sigma_1p_r +\sigma_2\frac{K}{R} +\sigma_3G\phi_0~~, 
		\label{H0.eqn3}\\
	&&H_1=-\sigma_2\frac{K}{R^2}w +\sigma_3G\phi_1~~, 
		\label{H1.eqn3}\\
	&&H_2=\sigma_2\frac{K}{R^3}w^2~~, \label{H2.eqn3}
\end{eqnarray}
in the power of $\delta_b/R$ (we need the expressions (\ref{Etot.eqn2}) to (\ref{Eb.eqn2}) in the power of $\delta_b/R$ up to the leading and the next-to-leading order). We also expand $E=\sum_iE_i$, $E_f=\sum_iE_f^{(i)}$, $E_b=\sum_iE_b^{(i)}$ and $\epsilon_f=\sum_i\epsilon_i$. We obtain Eqs.(\ref{Hfpsi.eqn2}) and (\ref{phiEqqq.eqn2}) for the fields in the leading order,
\begin{eqnarray}
H_0\psi_0 &=& \epsilon_0\psi_0~~, \label{psi0Eq.eqn3}\\
\frac{{\rm d}^2\phi_0}{{\rm d}w^2} &=& 
	\frac{\lambda}{2}\phi_0(\phi_0^2-v^2) +\frac{G}{4\pi R^2}
		\sum_{KM}\psi_0^{\dagger}\sigma_3\psi_0~~.
\label{phi0Eq.eqn3}
\end{eqnarray}

We first solve Eq.(\ref{psi0Eq.eqn3}). Taking into account that $\phi_0(\pm \infty)\rightarrow \pm v$, we obtain the normalizable solution,
\begin{equation}
\psi_0(w)=\frac{1}{\sqrt{{\cal N}}}e^{-U_0(w)}\chi_{+}~~,
\label{psi0.eqn3}
\end{equation}
with $U_0(w) = G\int_{0}^{w}{\rm d}w'~ \phi_0(w')$ and the normalization factor ${\cal N}=\int_{-\infty}^{+\infty}{\rm d}w~ e^{-2U_0(w)}$. Here,  $\chi_{\pm}$ are eigenspinors of $\sigma_2$ satisfying $\sigma_2\chi_{\pm}=\pm \chi_{\pm}$ and $\chi_{\pm}^{\dagger}\chi_{\pm}=1$.  We obtain the energy eigenvalue,
\begin{equation}
\epsilon_0=\frac{K}{R}~~,
\end{equation}
and take only $\epsilon_0$ positive, i.e., $\omega =1$. Using Eq.(\ref{psi0.eqn3}), we obtain the leading order fermi energy,
\begin{eqnarray}
E_f^{(0)} &=& \sum_{KM}\int_{-\infty}^{+\infty}{\rm d}w~ 
	\psi_0^{\dagger}H_0\psi_0 
	= 
	\sum_{KM}\frac{K}{R} 
	= 
	\frac{1}{R}\sum_{K=1}^{K_{max}} K(2K) \nonumber \\
	&=& 
	\frac{K_{max}(K_{max}+1)(2K_{max}+1)}{3R}~~.
\label{Ef0.eqn3}
\end{eqnarray}
Here, $K_{max}$ is determined by the total fermion number,
\begin{equation}
\sum_{KM} \int_{-\infty}^{+\infty}{\rm d}w~ \psi_0^{\dagger}\psi_0 
	= 
	\sum_{K=1}^{K_{max}} 2K 
	= 
	K_{max}(K_{max}+1) 
	= 
	N_f~~.
\label{Nf.eqn3}
\end{equation}
From Eqs.(\ref{Ef0.eqn3}) and (\ref{Nf.eqn3}), we get
\begin{equation}
E_f^{(0)}=\frac{2N_f^{\frac{3}{2}}}{3R} \left( 1+\frac{1}{4N_f} \right)^{\frac{1}{2}} 
	\simeq 
	\frac{2N_f^{\frac{3}{2}}}{3R}+\frac{N_f^{\frac{1}{2}}}{12R} \hspace{1.5cm} {\rm for}~~ N_f\gg 1~~.
\label{Ef00.eqn3}
\end{equation}
The first term is the leading contribution to the fermi energy, which is the same as that obtained within the exact thin-wall approximation \cite{MAC}. The second term in Eq.(\ref{Ef00.eqn3}) is the correction caused by the effect of quantumizing the angular momentum, which we call $\Delta E_f$ for the later discussion.

We next consider Eq.(\ref{phi0Eq.eqn3}). Since the leading solution for the fermion satisfies $\psi_0^{\dagger}\sigma_3\psi_0=0$, the equation (\ref{phi0Eq.eqn3}) becomes
\begin{equation}
\frac{{\rm d}^2\phi_0}{{\rm d}w^2} = 
	\frac{\lambda}{2}\phi_0(\phi_0^2-v^2)~~.
\label{phi0Eqq.eqn3}
\end{equation}
We know that the solution to the above equation satisfying $\phi_0(\pm \infty) \rightarrow \pm v$ is a kink,
\begin{equation}
\phi_0(w)=v\tanh{\frac{w}{\delta_b}}~~,
\label{phi0.eqn3}
\end{equation}
with $\delta_b=2/(\sqrt{\lambda}v)$. Using Eq.(\ref{phi0.eqn3}), we obtain the leading order surface energy,
\begin{equation}
E_b^{(0)}=4\pi R^2 \int_{-\infty}^{+\infty}{\rm d}w~ \left\{ 
	\frac{1}{2}\left( \frac{{\rm d}\phi_0}{{\rm d}w} \right)^2 
	+ \frac{\lambda}{8}(\phi_0^2-v^2)^2 
	\right\} 
	= 
	4\pi R^2 \Sigma~~,
\label{Eb0.eqn3}
\end{equation}
with $\Sigma=2\sqrt{\lambda}v^3/3$. This also coincides with the result obtained in Ref.\cite{MAC} within the thin-wall approximation.

Combining Eq.(\ref{Eb0.eqn3}) with the first term in Eq.(\ref{Ef00.eqn3}), we get the leading order Fermi Ball energy,
\begin{equation}
E_0 = \frac{2N_f^{\frac{3}{2}}}{3R} +4\pi R^2 \Sigma~~.
\label{E0.eqn3}
\end{equation}
( Note that we neglect the false vacuum volume energy, assuming that it is negligibly small. We discuss its magnitude in the cosmological context in Sec.\ref{concl.sec}.) Minimizing Eq.(\ref{E0.eqn3}) with respect to $R$, we get
\begin{equation}
R=\frac{\sqrt{N_f}}{(12\pi \Sigma)^{\frac{1}{3}}}
	= 
	\frac{\sqrt{N_f}}{(8\pi)^{\frac{1}{3}}\lambda^{\frac{1}{6}}v}~~,
\label{R.eqn3}
\end{equation}
which yields
\begin{equation}
E_0=(8\pi)^{\frac{1}{3}}\lambda^{\frac{1}{6}}N_fv~~.
\label{E00.eqn3}
\end{equation}
Here, let us examine the stability of the Fermi Ball against the fragmentation using the leading order Fermi Ball energy. We compare two states: the one in which a single Fermi Ball has the fermion number $N_f$ and the other in which $n$ Fermi Balls have the fermion numbers less than $N_f$ but keep the total fermion number to be $N_f$. Since the energy of the Fermi Ball in the leading order approximation is proportional to the total fermion number, the two states have the same energy. The leading order estimation cannot tell whether the Fermi Balls with large $N_f$ produced in the early universe survive until now or not \footnote{Taking into account the small energy density difference $\Lambda$ which we have neglected here, we see that the large Fermi Balls produced in the early universe have been fragmented into tiny pieces which are not thin-walled. The discussions on it are given in Ref.\cite{MAC,MOR}, and also in Ref.\cite{YOS} with some more details}. In order to examine the stability of the Fermi Ball, we calculate the higher order corrections in $E$ in the next section.

\section{HIGHER ORDER CORRECTION OF $\delta_b/R$-PERTURBATION\label{2ndorder.sec}}
The next-to-leading order components of the fields satisfy
\begin{eqnarray}
&& \left( H_0 -\frac{K}{R} \right)\psi_1 = -(H_1-\epsilon_1)\psi_0 
	=-H_1\psi_0~~, 
	\label{psi1Eq.eqn4}\\
&& \frac{{\rm d}^2\phi_1}{{\rm d}w^2} -\frac{\lambda}{2} 
	(3\phi_0^2-v^2)\phi_1 = -\frac{2}{R}\frac{{\rm d}\phi_0}{{\rm d}w} 
	+\frac{G}{4\pi R^2}\sum_{KM}(
		\psi_0^{\dagger}\sigma_3\psi_1 
		+\psi_1^{\dagger}\sigma_3\psi_0)~~.
	\label{phi1Eq.eqn4}
\end{eqnarray}
We here note
\begin{equation}
\epsilon_1=\int_{-\infty}^{+\infty}{\rm d}w~ 
	\psi_0^{\dagger}H_1\psi_0 =
	\frac{1}{{\cal N}}\int_{-\infty}^{+\infty}{\rm d}w~ we^{-2U_0(w)} 
	=0~~. 
\label{epsilon1.eqn4}
\end{equation}

\begin{figure}[htbp]
\includegraphics[height=16cm]{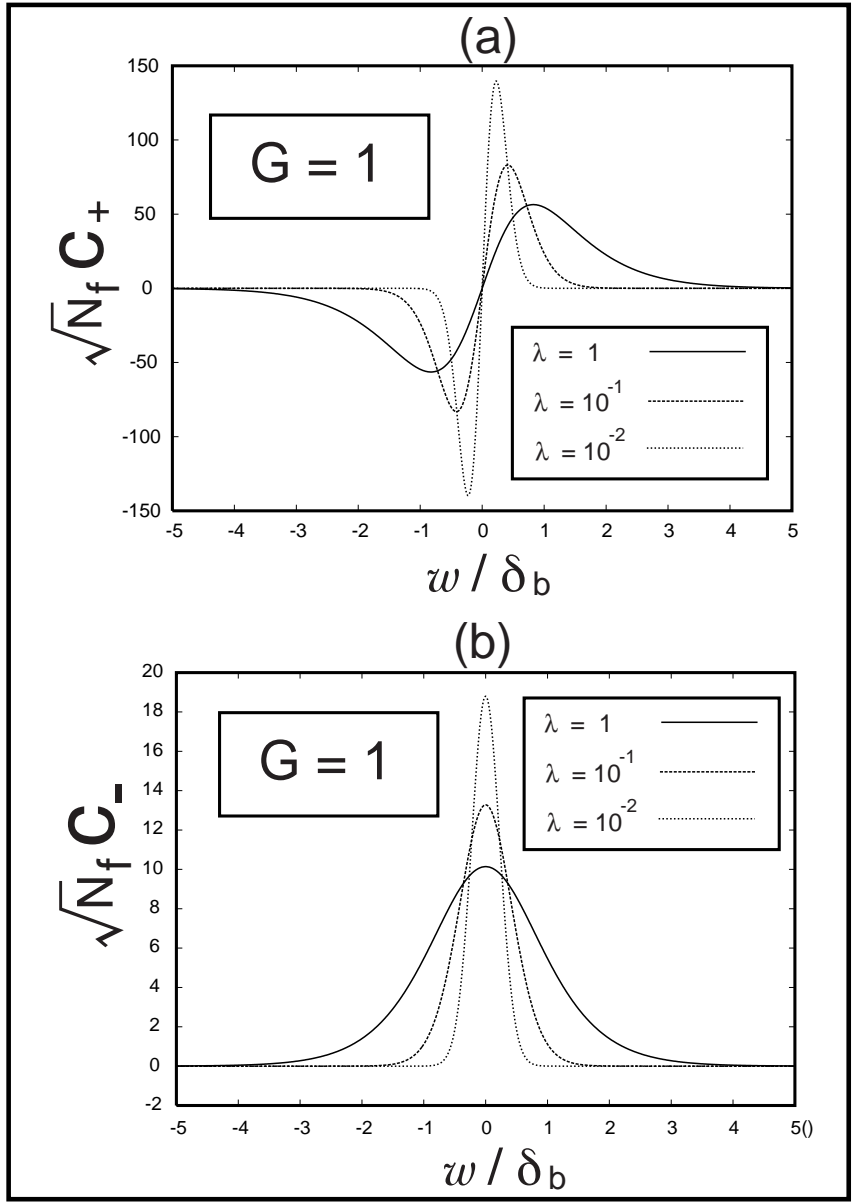}
\caption{The $\sigma_2=+1$ (Figure (a)) and $\sigma_2=-1$ (Figure (b)) component of the next-to-leading order solution for the fermion field with the Yukawa coupling constant $G=1$. In the figures, $c_{\pm}$ and $w$  are rescaled as $\sqrt{N_f}c_{\pm}$ and $w/\delta_b$, respectively, where $N_f$ is the total fermion number of the Fermi Ball and $\delta_b$ is the domain wall thickness. The origin of the $w$ axis denotes the center of the domain wall. Figure (a) shows that $c_{+}$ is getting more centrally localized around the origin as the scalar self-coupling constant $\lambda$ decreases.  Figure (b) shows that the width of $c_{-}$ around the origin decreases with the decreasing $\lambda$.\label{cplusminus.fig}}
\end{figure}

In order to solve Eq.(\ref{psi1Eq.eqn4}), we first write $\psi_1$ as the following form:
\begin{equation}
\psi_1(w)=\frac{1}{\sqrt{{\cal N}}} \left\{ 
	\xi_1(w)\psi^{(a)}(w) +\xi_2(w)\psi^{(b)}(w)
	\right\}~~,
\label{psi1.eqn4}
\end{equation}
where $\psi^{(a)}(w)$ and $\psi^{(b)}(w)$ are the linearly independent solutions to Eq.(\ref{psi0Eq.eqn3}) with the same eigenvalue $\epsilon_0$, \begin{eqnarray}
\psi^{(a)}(w)&=&e^{-U_0(w)}\chi_{+}
\label{psia.eqn4} \\
\psi^{(b)}(w) &=& \frac{K}{R}e^{-U_0(w)}W(w)\chi_{+} 
	+\frac{1}{2}e^{U_0(w)}\chi_{-}~~, \label{psib.eqn4}
\end{eqnarray}
with $W(w) = \int_{0}^{w}{\rm d}w'~ e^{2U_0(w')}$. The equations (\ref{psi1Eq.eqn4}) and (\ref{psi1.eqn4}) give
\begin{eqnarray}
&&\frac{{\rm d}\xi_1}{{\rm d}w} 
	+\frac{K}{R}W(w)\frac{{\rm d}\xi_2}{{\rm d}w} =0~~, 
\label{xi12Eq1.eqn4}\\
&&\frac{{\rm d}\xi_2}{{\rm d}w} 
	+\frac{2K}{R^2}we^{-2U_0(w)} =0~~.
\label{xi12Eq2.eqn4}
\end{eqnarray}
The solutions to the above equations are
\begin{eqnarray}
\xi_1(w) &=& \frac{2K^2}{R^3} \int_{0}^{w}{\rm d}w'~ 
	w'W(w')e^{-2U_0(w')} -U_1(w)~~, \label{xi1.eqn4}\\
\xi_2(w) &=& \frac{2K}{R^2} \int_{w}^{+\infty}{\rm d}w'~ 
	w'e^{-2U_0(w')}~~, \label{xi2.eqn4}
\end{eqnarray}
with $U_1(w)=G\int_{0}^{w}{\rm d}w' \phi_1(w')$. We thus obtain the solution for $\psi_1$,
\begin{equation}
\psi_1(w)=\frac{1}{\sqrt{{\cal N}}}
	\biggl(
	c_{+}(w)\chi_{+}+c_{-}(w)\chi_{-}
	\biggr)~~,
\label{psi11.eqn4}
\end{equation}
where
\begin{eqnarray}
c_{+}(w)&=& e^{-U_0}
	\Biggl(
	\frac{2K^2}{R^3}\int_0^w{\rm d}w'~ e^{2U_0}
	\int_{w'}^{\infty}{\rm d}w''~ 
	w''e^{-2U_0}
	-U_1(w)
	\Biggr)~~,
\label{cp.eqn4}\\
c_{-}(w)&=& \frac{K}{R^2}e^{U_0}
	\int_w^{+\infty}{\rm d}w'~w'e^{-2U_0}~~.
\label{cm.eqn4}
\end{eqnarray}
The $w$-dependence of the component $c_{\pm}$ is shown in FIG.\ref{cplusminus.fig}.

\begin{figure}[htbp]
\includegraphics[height=10cm]{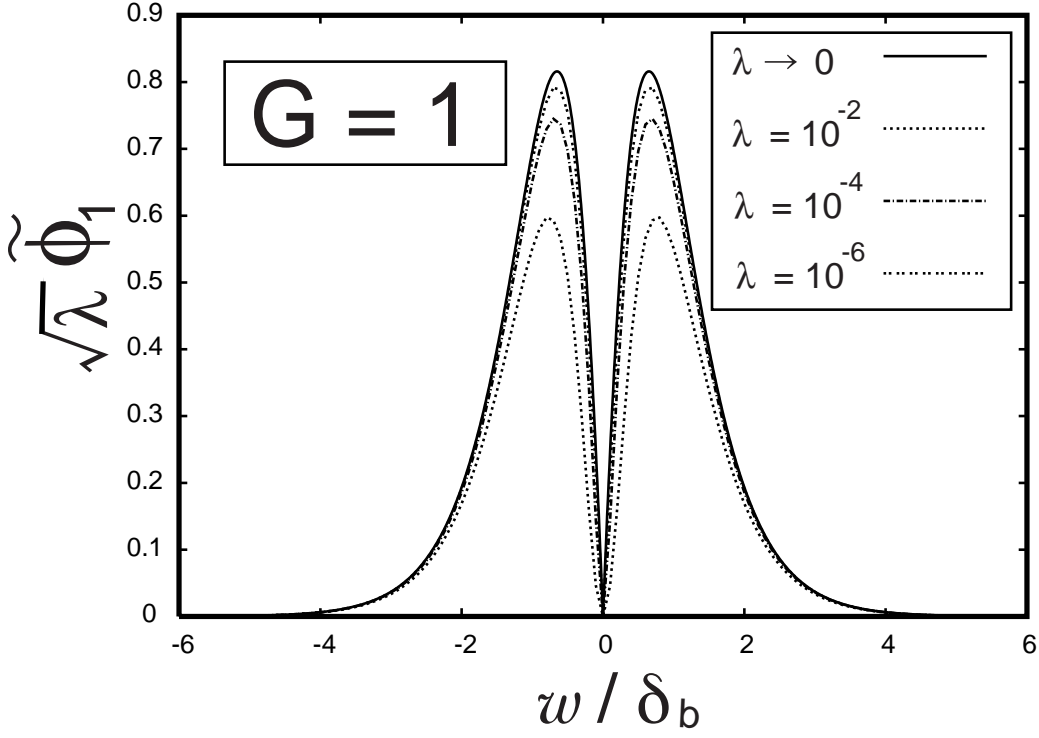}
\caption{The next-to-leading order solution $\phi_1$ for the domain wall field $\phi$. In the figure, $\phi_1$ and $w$ are rescaled as $\sqrt{\lambda}\widetilde{\phi_1}\equiv \sqrt{\lambda}R\phi_1$ and $w/\delta_b$, respectively. Here, $\lambda$ is the scalar self-coupling constant and $R$ and $\delta_b$ are the Fermi Ball radius and the wall thickness, respectively. The origin of the $w$ axis denotes the center of the domain wall. The figure shows that $\phi_1$ is not smooth at the origin in the limit of $\lambda \rightarrow 0$, which coincides with the case where the spreading width of the fermionic source term in Eq.(\ref{phi1Eq.eqn4}) is neglected (see Ref.\cite{YOS}). \label{phi1.fig}}
\end{figure}

We next solve the Eq.(\ref{phi1Eq.eqn4}). Replacing the fermionic source term of Eq.(\ref{phi1Eq.eqn4}) by $\psi_0^{\dagger}\sigma_3\psi_1=\psi_1^{\dagger}\sigma_3\psi_0=\xi_2/(2{\cal N})$, we get
\begin{eqnarray}
\left[ \frac{{\rm d}^2}{{\rm d}w^2} -\frac{\lambda}{2}(3\phi_0^2-v^2) 
	\right]\widetilde{\phi_1} 
	&=& 
	-2\frac{{\rm d}\phi_0}{{\rm d}w} +\frac{G}{4\pi R{\cal N}}
		\sum_{KM}\xi_2 \nonumber \\
	&\equiv& h(w)~~, \label{tphi1Eq.eqn4}
\end{eqnarray}
where $\widetilde{\phi_1}=R\phi_1$. The solution to the above equation is given by
\begin{eqnarray}
\widetilde{\phi_1}(w)&=& \frac{1}{{\rm cosh}^2\frac{w}{\delta_b}}
	\int_{0}^{w}{\rm d}w'~ {\rm cosh}^4\frac{w'}{\delta_b}
	\left\{
	\int_{0}^{w'}-\int_{0}^{\infty}
	\right\}
	{\rm d}w''~ \frac{h(w'')}{{\rm cosh}^2\frac{w''}{\delta_b}}
	\nonumber\\
	&=& \frac{1}{{\rm cosh}^2\frac{w}{\delta_b}}
	\int_{0}^{w}{\rm d}w'~ {\rm cosh}^4\frac{w'}{\delta_b}
	\int_{0}^{w'}{\rm d}w''~ 
		\frac{h(w'')}{{\rm cosh}^2\frac{w''}{\delta_b}}
\label{tphi1.eqn4}
\end{eqnarray}
where we use
\begin{equation}
\int_{0}^{\infty}{\rm d}w~
	\frac{h(w)}{{\rm cosh}^2\frac{w}{\delta_b}}
	\propto
\int_{0}^{\infty}{\rm d}w~ \frac{{\rm d}\phi_0}{{\rm d}w}h(w)
=0
\label{0hint.eqn4}
\end{equation}
Note that $\widetilde{\phi_1}(w)$ satisfies $\widetilde{\phi_1}(0)=0$ and $\widetilde{\phi_1}(\pm \infty) \rightarrow 0$ (see FIG.\ref{phi1.fig}). 

\begin{figure}[htbp]
\includegraphics[height=10cm]{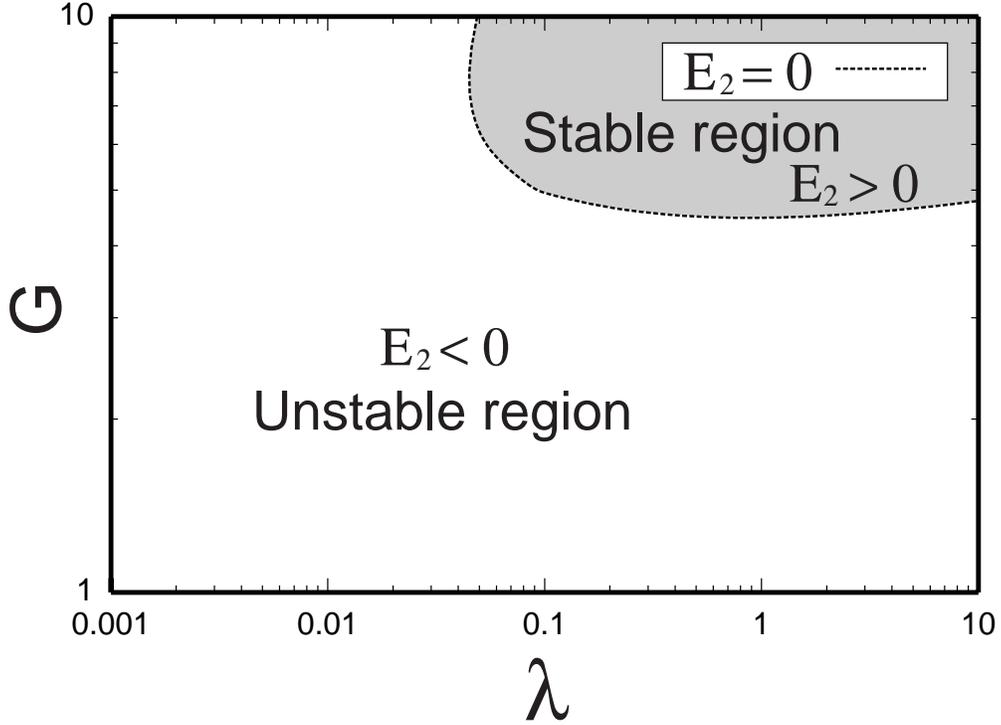}
\caption{The region of the scalar self-coupling constant $\lambda$ and the Yukawa coupling constant $G$ where the neutral Fermi Ball is stable (shadowed region) or unstable (blank region) against the fragmentation. The energy $E_2$ in the figure denotes the deviation from the leading order energy of the Fermi Ball obtained within the thin-wall analysis. We see that rather large value of $G$ is allowed for the Fermi Ball to be stable (see also FIG.\ref{onphase.fig} and its caption).\label{phase.fig}}
\end{figure}

\begin{figure}[htbp]
\includegraphics[height=10cm]{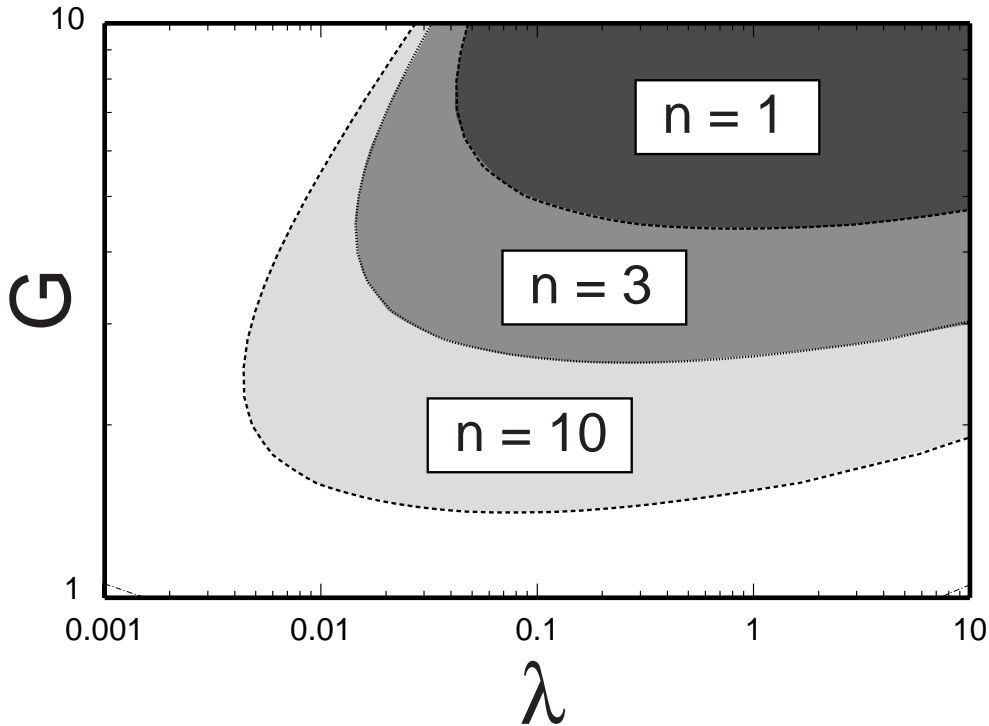}
\caption{The allowed region (shadowed) of the scalar self-coupling constant $\lambda$ and the Yukawa coupling constant $G$ for the Fermi Ball with multi-fermions $\Psi_i$ ($1\leq i\leq n$) to be stable against the fragmentation (see text). The allowed region is illustrated for $n=1,3~{\rm and}~10$. The figure shows that the region extends as $n$ is increases. \label{onphase.fig}}
\end{figure}

We note that the first order energy corrections $E_f^{(1)}$ and $E_b^{(1)}$ vanish since the integrands for them are odd functions of $w$. In the next-to-leading order, the fermi energy is
\begin{eqnarray}
E_f^{(2)} &=& 
	\sum_{KM}\int_{-\infty}^{+\infty}{\rm d}w~\psi_0^{\dagger}H_2\psi_0
	+\sum_{KM}\int_{-\infty}^{+\infty}{\rm d}w~\psi_0^{\dagger}H_1\psi_1
	\nonumber\\
	&\simeq&
\frac{2N_f^{\frac{3}{2}}}{3R^3{\cal N}}\int_{-\infty}^{+\infty}{\rm d}w~
	w^2e^{-2U_0} 
+\frac{4GN_f^{\frac{3}{2}}}{3R^3{\cal N}}\int_{-\infty}^{+\infty}{\rm d}w~
	we^{-2U_0}\int_{0}^{w}{\rm d}w'~
		\widetilde{\phi_1}(w')	\nonumber\\
&&
-\frac{4N_f^{\frac{5}{2}}}{5R^5{\cal N}}\int_{-\infty}^{+\infty}{\rm d}w~
	we^{-2U_0}\int_{0}^{w}{\rm d}w'~
		e^{2U_0}\int_{w'}^{+\infty}{\rm d}w''~
			w''e^{-2U_0}
\hspace{0.7cm} {\rm for}~~ N_f\gg 1 \nonumber\\
&=&
\frac{16\pi \lambda^{\frac{1}{2}}v^3}{3{\cal N}}
\int_{-\infty}^{+\infty}{\rm d}w~
	w^2e^{-2U_0} 
+\frac{32\pi \lambda^{\frac{1}{2}}Gv^3}{3{\cal N}}
\int_{-\infty}^{+\infty}{\rm d}w~
	we^{-2U_0}\int_{0}^{w}{\rm d}w'~
		\widetilde{\phi_1}(w')	\nonumber\\
&&
-\frac{128\pi^{\frac{5}{3}}\lambda^{\frac{5}{6}}v^5}{5{\cal N}}
\int_{-\infty}^{+\infty}{\rm d}w~
	we^{-2U_0}\int_{0}^{w}{\rm d}w'~
		e^{2U_0}\int_{w'}^{+\infty}{\rm d}w''~
			w''e^{-2U_0}~~,
\label{Ef2.eqn4}
\end{eqnarray}
and the surface energy is
\begin{eqnarray}
E_b^{(2)} &=& 
4\pi R^2\int_{-\infty}^{+\infty}{\rm d}w~
	\frac{w^2}{R^2}\left\{
		\frac{1}{2}\left(
			\frac{{\rm d}\phi_0}{{\rm d}w}
		\right)^2
		+\frac{\lambda}{8}(\phi_0^2-v^2)^2
	\right\} \nonumber\\
&&
+4\pi R^2\int_{-\infty}^{+\infty}{\rm d}w~
	\frac{2w}{R}\left\{
		\frac{{\rm d}\phi_0}{{\rm d}w}\frac{{\rm d}\phi_1}{{\rm d}w}
		+\frac{\lambda}{2}\phi_0\phi_1(\phi_0^2-v^2)
	\right\} \nonumber\\
&&
+4\pi R^2\int_{-\infty}^{+\infty}{\rm d}w~
	\left\{
		\frac{1}{2}\left(
			\frac{{\rm d}\phi_1}{{\rm d}w}
		\right)^2
		+\frac{\lambda}{4}\phi_1^2(3\phi_0^2-v^2)
	\right\} \nonumber\\
&=&
4\pi \int_{-\infty}^{+\infty}{\rm d}w~
	\left\{
		w^2 \left(\frac{{\rm d}\phi_0}{{\rm d}w} \right)^2
		-2\widetilde{\phi_1}\frac{{\rm d}\phi_0}{{\rm d}w} 
		-\frac{1}{2}\widetilde{\phi_1}\frac{h(w)}{R} 
	\right\} \nonumber\\
&=&
4\pi \int_{-\infty}^{+\infty}{\rm d}w~
	\left\{
		w^2 \left(\frac{{\rm d}\phi_0}{{\rm d}w} \right)^2
		-\widetilde{\phi_1}\frac{{\rm d}\phi_0}{{\rm d}w}
	\right\}
-\frac{G}{2R{\cal N}}\sum_{KM}\int_{-\infty}^{+\infty}{\rm d}w~
	\widetilde{\phi_1}\xi_2 \nonumber\\
&=&
\pi\lambda v^4 \int_{-\infty}^{+\infty}{\rm d}w~
	\frac{w^2}{{\rm cosh}^4\frac{w}{\delta_b}}
-2\pi\lambda^{\frac{1}{2}} v^2 \int_{-\infty}^{+\infty}{\rm d}w~
	\frac{\widetilde{\phi_1}(w)}{{\rm cosh}^2\frac{w}{\delta_b}}
\nonumber\\
&& \hspace{1cm}
-\frac{16\pi\lambda^{\frac{1}{2}}Gv^3}{3{\cal N}}
\int_{-\infty}^{+\infty}{\rm d}w~
	we^{-2U_0} \int_{0}^{w}{\rm d}w'~
		\widetilde{\phi_1}(w')~~,
\label{Eb2.eqn4}
\end{eqnarray}
where $R$ is replaced by Eq.(\ref{R.eqn3}). Taking into account Eq.(\ref{R.eqn3}) and $\Delta E_f$, the second term in the r.h.s. of Eq.(\ref{Ef00.eqn3}), we obtain the Fermi Ball energy in the next-to-leading order, 
\begin{equation}
E_2=E_f^{(2)}+E_b^{(2)}+\Delta E_f={\cal C}(\lambda,G)v~~. 
\label{E2final.eqn4}
\end{equation}
We see that ${\cal C}(\lambda, G)$ does not depend on $N_f$; this is crucial to the discussion on the stability of the Fermi Ball below. Let us consider a single-Fermi Ball state and a $n$-Fermi Balls state both with   the total fermion number taken to be $N_f$. The two states have the same leading-order energies $E_0$, while they have the different higher-order corrections $E_2$: the former has a correction ${\cal C}v$ and the latter $n{\cal C}v$. Therefore, when ${\cal C}$ is positive, the former state has the lower energy and the Fermi Ball is stable against the fragmentation. We evaluate $E_2$ numerically and find that ${\cal C}(\lambda,G)$ is positive in a limited parameter region of $\lambda$ and $G$ (see FIG.\ref{phase.fig}). 
We see in FIG.\ref{phase.fig} that rather large value of $G$ is allowed for the Fermi Ball to be stable. This situation, however, is much changed if the fermions have more degrees of freedom, e.g., belonging to a large multiplet of the internal symmetry and if the scalar field $\phi$ belongs to a singlet (see the forthcoming paper). Consider the fermion multiplet $\Psi_i$ ($1\leq i\leq n$) coupling to $\phi$ through the common Yukawa coupling constant $G$, and assume for simplicity the fermion number $N_i=N$ in common for each flavor $i$. In such a case, we see that the energy correction $E_2$ is independent of $N$ but depends on $n$, and that a smaller value of $G$ can stabilize the Fermi Ball (see FIG.\ref{onphase.fig}). Here, we emphasize that there is a certain region of the parameters where stable Fermi Balls are allowed to exist. 

\section{Cosmological or observational constraints on region of parameters}
\label{para.sec}
We very roughly examine here cosmological or observational constraints on the parameter region for the neutral Fermi ball.  We define the parameter $\kappa = 2\pi^{1/3}\lambda^{1/6}v$ which satisfies
\begin{equation}
M_f = \kappa^3 R^2~~,
\label{kappa_neu.eqn4}
\end{equation}
with the Fermi ball mass $M_f$ and the radius $R$ being given by Eq.(\ref{E00.eqn3}) and Eq.(\ref{R.eqn3}), respectively.

We first consider the energy density difference $\Lambda$ which we have neglected so far. We know that $\Lambda$ should be larger than the critical value $\Lambda_c \simeq \kappa^6/(144\pi^2 M_{pl}^2)$ in order to avoid the domination of the black holes made up by domain walls in the total energy density of the early universe \cite{VILL}. Such a finite $\Lambda$ gives a volume energy $E_v= 4\pi \Lambda R^3/3$ to the false vacuum, having an effect to destabilize the Fermi Ball. As far as the volume energy is small enough, i.e., $E_v = 4\pi \Lambda M_f^{3/2}/(3\kappa^{9/2})<E_2$, the Fermi Ball is absolutely stable. This gives a constraint, 
\begin{equation}
M_f < 2.9\times 10^{26} \left(\frac{{\rm GeV}}{\kappa}\right)^{\frac{1}{3}}
\left(\frac{E_2}{\kappa}\right)^{\frac{2}{3}}~{\rm GeV}~~,
\label{stable_neu.eqn4}
\end{equation}
due to the condition for $\Lambda$ to exist under the above constraints. 

We next consider the observational aspect of the neutral Fermi ball.  If they are produced in the early universe  and have survived until present, they can contribute to the dark matter in the Galaxy. Let us assume here that the neutral stable Fermi ball has a sizable contribution to dark matter. We then have their flux $F$,
\begin{equation}
F 
\simeq \frac{\rho_{DM}~ u_0}{4 \pi M_f} 
\sim 
7.1 \times 10^{5} \left ( \frac{\mbox{GeV}}{M_f} \right) 
~{\rm cm^{-2}sec^{-1}sr^{-1}}~,
\label{DM-limit.eqn4}
\end{equation}
where $\rho_{DM}\sim 0.3 ~{\rm GeV~cm^{-3}}$ is the energy density of the dark matter in the Galaxy and $u_0 \sim 3 \times 10^7 ~{\rm cm~sec^{-1}}$ is the Virial velocity of the Fermi ball.  We seek for the allowed region of the Fermi ball parameters through the use of currently available observational data. We use in the following the results of the experiments which searched for monopoles or heavy dark matter.

In order to examine whether the Fermi ball can be detected in terrestrial experiments, we first consider the condition that the neutral Fermi ball should reach the detector passing through matter. The energy loss rate (energy loss per path length) is given by \cite{Wit,Gla}
\begin{equation}
\frac{{\rm d}E}{{\rm d}x} = -  \sigma u^2\rho~~ ,
\label{eloss.eqn4}
\end{equation}
where $\sigma$ is the collision cross section with a nucleus in the medium, $u$ is the velocity of the Fermi ball, and $\rho$ is the density of the target matter. For simplicity, in the following we assume that the cross section is geometrically given by,
\begin{equation}
\sigma = \pi R^2~~.
\label{cross_neu.eqn4}
\end{equation}
From Eq.(\ref{eloss.eqn4}), the velocity $u$ decreases exponentially with the path length, $\propto {\rm exp}(-\sigma \rho x/M_f)$. Estimating the final velocity as $u_c\sim 1.2\times 10^4~{\rm cm~sec^{-1}}$ \cite{Gla}, we obtain the condition for the Fermi ball to reach the detector,
\begin{equation}
M_f > 0.13\sigma \rho L~~,
\label{reach.eqn4}
\end{equation}
where $L$ is the path length. Substituting Eqs.(\ref{kappa_neu.eqn4}) and (\ref{cross_neu.eqn4}) into Eq.(\ref{reach.eqn4}) yields
\begin{equation}
\kappa >
4.6\times 10^{-2}\left(\frac{\rho L}{{\rm gr~cm^{-2}}}\right)^{1/3}
~{\rm GeV}~~.
\label{reach_neu.eqn4}
\end{equation}

We next consider the efficiency of the detectors to observe the neutral Fermi ball. Let us examine the track detectors with mica \cite{Mic} and the scintillators in the MACRO \cite{Mcr} and the KEK \cite{Kek} experiments. These experiments are sensitive to the neutral Fermi ball, if the energy loss per path length is large enough, $q({\rm d}E/\rho {\rm d}x) > ({\rm d}E/\rho {\rm d}x)_{min}$, where $({\rm d}E/\rho {\rm d}x)_{min}$ is the detection threshold for relativistic charged particles and $q$ is the efficiency correction factor. This condition with Eq.(\ref{eloss.eqn4}) gives $\sigma > 1.7\times 10^{-18}[q^{-1}({\rm d}E/\rho {\rm d}x)_{min}/{\rm GeV/gr~cm^{-2}}]~{\rm cm}^2$, that is,
\begin{equation}
M_f=\frac{\sigma \kappa^3}{\pi}
>1.4\times 10^9 \left(\frac{\kappa}{{\rm GeV}}\right)^3
\left[\frac{q^{-1}({\rm d}E/\rho {\rm d}x)_{min}}{{\rm GeV/gr~cm^{-2}}}\right]~{\rm GeV}~~.
\label{efficient.eqn4}
\end{equation}
For the scintillator experiments in MACRO and KEK we take $q=1$  for simplicity \footnote{The value $q$ in this case is called a quenching factor and ranges $q=0.1\sim 1$ with various conditions. We are interested only in the order of magnitude estimation, and we take $q=1$ in this paper.}. In case of the MACRO experiments, the conditions Eq.(\ref{reach_neu.eqn4}) and Eq.(\ref{efficient.eqn4}) with $\rho L=3.7\times 10^{5}~{\rm gr~cm^{-2}}$ and $q^{-1}({\rm d}E/\rho {\rm d}x)_{min}\sim 3I_{min}\sim 6~{\rm MeV/gr~cm^{-2}}$ \cite{Mcr} give $\kappa > 3.3~{\rm GeV}$ and $M_f > 7.8\times 10^{6}(\kappa/{\rm GeV})^3~{\rm GeV}$. The flux upper limit of $F < 2.5\times 10^{-16}~{\rm cm^{-2}sec^{-1}sr^{-1}}$ with Eq.(\ref{DM-limit.eqn4}) gives the mass lower limit $M_f > 2.8\times 10^{21}~{\rm GeV}$. In case of the KEK experiments, the conditions $\rho L=10^{3}~{\rm gr~cm^{-2}}$ and  $q^{-1}({\rm d}E/\rho {\rm d}x)_{min}\sim 0.01I_{min}\sim 20~ {\rm KeV/gr~cm^{-2}}$ \cite{Mcr} give $\kappa > 0.46~{\rm GeV}$ and $M_f > 2.6\times 10^{4}(\kappa/{\rm GeV})^3~{\rm GeV}$. The flux bound $F < 3.2\times 10^{-11}~{\rm cm^{-2}sec^{-1}sr^{-1}}$ gives $M_f > 2.2\times 10^{16}~{\rm GeV}$.

For the truck detector with mica \cite{Mic} we also take $q=1$ for simplicity. In this case we take $\rho L=7.5\times 10^{5}~{\rm gr~cm^{-2}}$ which is due to the fact that the mica was located at 3 km deep under the earth, and $q^{-1}({\rm d}E/\rho {\rm d}x)_{min}\sim 2.4~{\rm GeV/gr~cm^{-2}}$  \cite{Mic}. They give $\kappa > 4.2~{\rm GeV}$ and $M_f > 3.1\times 10^{9}(\kappa/{\rm GeV})^3~{\rm GeV}$. The flux upper limit $F < 2.3\times 10^{-20}~{\rm cm^{-2}sec^{-1}sr^{-1}}$ gives $M_f > 3.1\times 10^{25}~{\rm GeV}$.

Recently very low background experiments are being done for the dark matter search \cite{Cdms,Dam,Min}. Here, we analyze the results of CDMS experiments, which uses cryogenic Ge detector. The condition Eq.(\ref{reach_neu.eqn4}) with $\rho L=2.6\times 10^{3}~{\rm gr~cm^{-2}}$ gives $\kappa > 1.2~{\rm GeV}$. The dark matter search experiments allow such a small cross section of the cold dark matter particle scattering with the target nucleus that the average number of collisions in the target is less than a unity ($\sigma \rho d/m_{Ge} \sim \sigma \rho d/A_{Ge}m_p < 1$, namely, $\sigma < 2.2\times 10^{-23}~{\rm cm^2}$ with $\rho = 5.3~{\rm gr/cm^3}$, $d=1cm$ and $A_{Ge} \sim 73$). The experimental results are roughly expressed as $F\sigma \propto \sigma/M_f < 3\times A_{Ge}\times 10^{-42}~{\rm cm^2/GeV}$ for $M_f > 100~{\rm GeV}$, and this gives $\kappa > 3.9\times 10^{4}~{\rm GeV}$.

\begin{figure}[htbp]
\centering
\includegraphics[height=8cm]{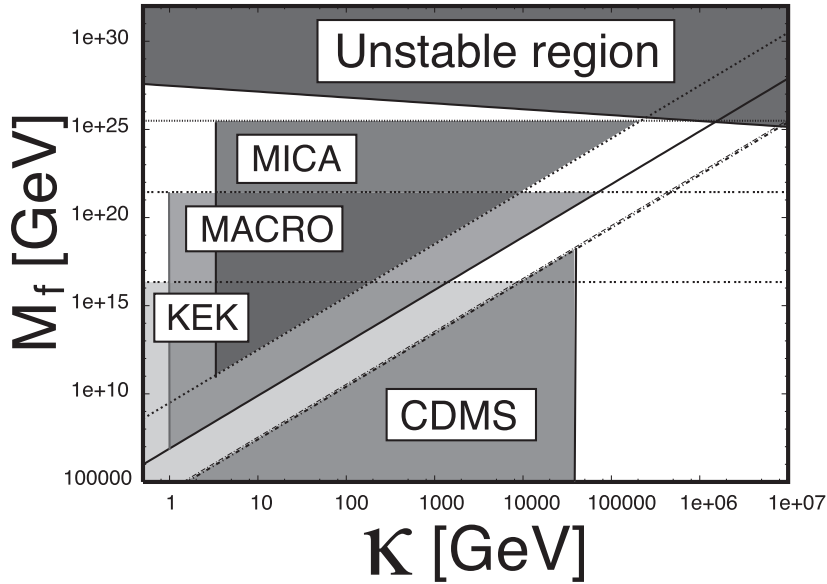}
\caption{The allowed regions (blank) of the Fermi ball mass $M_f$ and the quantity $\kappa$ defined by $\kappa = 2\pi^{1/3}\lambda^{1/6}v$ where $\lambda$ is the scalar self-coupling constant and $v$ is the symmetry breaking scale. The upper shadowed region is excluded by the stability condition. The other shadowed regions are excluded by the experiments MACRO \cite{Mcr}, KEK \cite{Kek}, MICA \cite{Mic} and CDMS \cite{Cdms}. We take $q=1$ for simplicity (see text). \label{exp.fig}}
\end{figure}

We illustrate the allowed region of $M_f$ and $\kappa$ in Figure \ref{exp.fig}. 
In this figure the condition for the stability Eq.(\ref{stable_neu.eqn4}) is also shown. One can see that there still remains the parameter region to be explored, especially the region of $\kappa \gtrsim 10^5~{\rm GeV}$ for $M_f \lesssim 10^{25}~{\rm GeV}$ and $\kappa \lesssim 10^5~{\rm GeV}$ for $10^{25}~{\rm GeV} \lesssim M_f \lesssim 10^{27}~{\rm GeV}$.

\section{CONCLUSION AND DISCUSSION\label{concl.sec}}
We have considered the neutral Fermi Balls in the thin-wall model where the domain wall thickness $\delta_b$ is much smaller than the Fermi Ball radius $R$. In the case where the spreading thickness $\delta_f$ of the fermion confined in the domain wall is negligibly small compared to $\delta_b$, the Fermi Ball is unstable against the fragmentation even if the finite $\delta_b$ is taken into account \cite{YOS}.  In the present paper, we have examined whether the Fermi Ball is stable or not if the effect of the finite $\delta_f$ is included.

In order to estimate the energy of the Fermi Ball, we have expanded the fields and the Hamiltonian in the power of $\delta_b/R$. At each level of the perturbation, we have examined the stability against the fragmentation. We have found that the energy correction in the next-to-leading order can stabilize the Fermi Ball in the limited region of the scalar self-coupling constant $\lambda$ and the Yukawa coupling constant $G$, as is shown in FIG.\ref{phase.fig}.

We have lastly given rough estimations for the allowed region of the parameters, $M_f$ and $\kappa$, for the neutral Fermi Ball to have a sizable contribution to the cold dark matter in case where the cross section of the Fermi Ball scattering with matter is of geometrical size. We have found that the allowed region is severely restricted by cosmological or observational constraints, but that there still remains the region open to the future exploration.

In the present paper, we have dealt with the Fermi Ball in a semi-classical manner. The quantum corrections, such as the radiative correction coming from the scattering of the fermions in the domain wall, may affect the stability of the Fermi Ball. They will be discussed elsewhere.



\begin{thebibliography}{999}
\bibitem{KIB}T. W. B. Kibble, J. Phys. {\bf A9} (1976) 1387.
\bibitem{VIL}A. Vilenkin, Phys. Rep. {\bf 121} (1985) 263.
\bibitem{GEL}G. B. Gelmini, M. Gleiser and E. W. Kolb, Phys. Rev. {\bf D39} (1989) 1558.
\bibitem{DVA}G. Dvali and M. Shifman, Nucl. Phys. {\bf B504} (1997) 127.
\bibitem{SAK}M. Sakamoto and M. Tachibana, Phys. Lett. {\bf B458} (1999) 231.
\bibitem{MAC}A. L. Macpherson and B. A. Campbell, Phys. Lett. {\bf B347}
(1995) 205.
\bibitem{MACC}A. L. Macpherson and J. L. Pinfold, hep-ph/9412264.
\bibitem{BRA}R. Brandenberger, I. Halperin and A. Zhitnitsky, hep-ph/9903318.
\bibitem{YOS}T. Yoshida, K. Ogure and J. Arafune, hep-ph/0210062.
\bibitem{MOR}J. R. Morris, Phys. Rev. {\bf D59} (1999) 023513.
\bibitem{VILL}A. Vilenkin and E. P. S. Shellard, {\it Cosmic String and Other Topological Defects} (Cambridge University Press, Cambridge, 1994).
\bibitem{Wit}E. Witten, Phys. Rev. {\bf D30} (1984) 272.
\bibitem{Gla}A. De R\'{u}jula and S. L. Glashow, {\it NATURE} {\bf
  312} (1984) 734.
\bibitem{Oya}S. Orito et al., Phys. Rev. Lett. {\bf 66} (1991) 1951.
\bibitem{Mic}P. B. Price and M. H. Salamon, Phys. Rev. Lett. {\bf 56}
  (1986) 1226.
\bibitem{Cdms}R. Abusaidi et al., Phys. Rev. Lett. {\bf 84} (2000) 5699.
\bibitem{Mcr}The MACRO Collaboration, Eur. Phys. J. {\bf C25} (2002) 511.
\bibitem{Bak}D. Bakari et al., Astropart. Phys. {\bf 15} (2001) 137.
\bibitem{Kek}K. Nakamura et al., Phys. Lett. {\bf B161} (1985) 417.
\bibitem{Ruj}A. De R\'{u}jula, S. L. Glashow, and Uri Sarid,
  Nucl. Phys. {\bf B333} (1990) 173.
\bibitem{Dam}R. Bernabei, et. al., Eur. Phys. J. {\bf C23} (2002) 61.
\bibitem{Min}Y. Shimizu, M. Minowa, H. Sekiya and Y. Inoue, astro-ph/0207529, accepted for publication in NIM-A.
\end{thebibliography}
\end{document}